\AtBeginDocument{\def\draftnote{July 10, 2013}}

\documentclass{ws-ijmpcs}
\usepackage{psfrag,units,wrapfig}

\newcommand{\VC}[1]{%
  \begin{tabular}[c]{l}%
    #1%
  \end{tabular}
}

\DeclareRobustCommand*\diff[2][]{%
   \mathop{
     \mathrm{d}^{#1}
     \mskip-0.2\thinmuskip
   #2}\nolimits
}

\newcommand\ifrac[2]{#1/#2}

\let\oldleft\left
\def\xleft{\mathopen{}\oldleft}

\newcommand{\T}[1]{\boldsymbol{#1}_{\text{T}}}
\newcommand{\Tj}[2]{\boldsymbol{#1}_{#2\,\text{T}}}
\def\Tsub#1_#2{\Tj{#1}{#2}}
\newcommand{\Tsc}[1]{#1_{\text{T}}}
\newcommand{\Tscj}[2]{#1_{#2\,\text{T}}}

\newcommand{\eqbad}{\stackrel{\textrm{?}}{=}}

\newcommand\bstar{\boldsymbol{b}_*}
\newcommand\bstarsc{b_*}
\newcommand\mub{\mu_b}
\newcommand\muQ{\mu_Q}

\begin{document}

\markboth{Collins}{TMD theory, factorization and evolution}

%
\catchline{}{}{}{}{}
%

\title{TMD THEORY, FACTORIZATION AND EVOLUTION}

\author{JOHN COLLINS}
\address{Physics Department, Penn State University, 
    University Park PA 16802, U.S.A.
\\
   collins@phys.psu.edu
}

\maketitle

\begin{history}
\end{history}

\begin{abstract}
  The concepts and methods of factorization using
  transverse-momentum-dependent (TMD) parton densities and/or
  fragmentation functions are summarized.
\end{abstract}


\section{Introduction}

TMD factorization is hard-scattering factorization in which
transverse-momentum-dependent (TMD) parton densities and/or
fragmentation functions are used.  This contribution summarizes the
theory of TMD factorization.  After a discussion of the physical
issues, I state TMD factorization and its main properties in the case
of the Drell-Yan (DY) process.  Then I explain what non-perturbative
information is involved, and summarize the predictive power of the
formalism.  The Collins-Soper-Sterman (CSS)
formalism\cite{Collins:1981uk,Collins:1984kg} is used, in the updated
form given in Ref.\ \refcite{Collins:2011qcdbook}.  It encodes
properties of QCD, so other valid formalisms must contain comparable
physics.

\section{Basic parton model inspiration: Case of Drell-Yan at
  $\Tsc{q} \ll Q$}

\begin{figure}
  \centering
  \begin{tabular}{c@{\hspace*{12mm}}c}
    \includegraphics[scale=0.4]{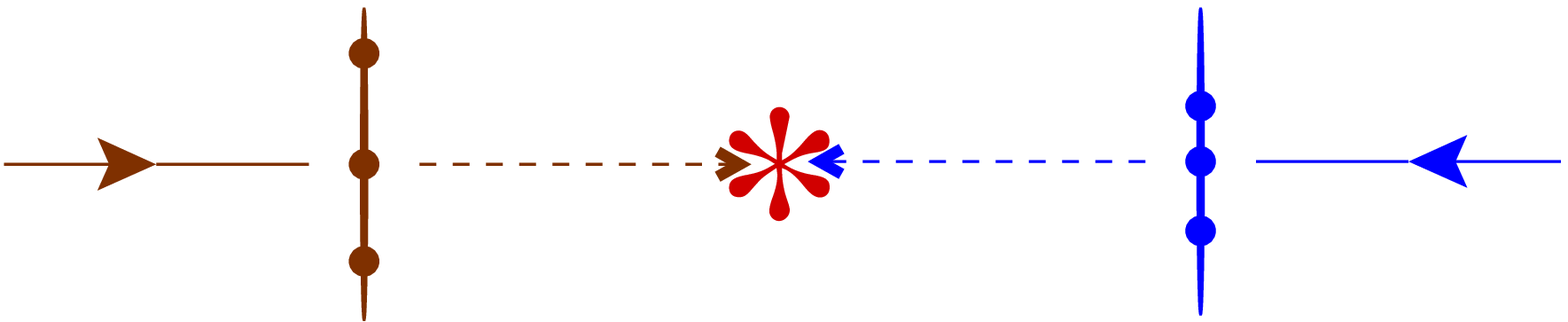}
    &
    \includegraphics[scale=0.55]{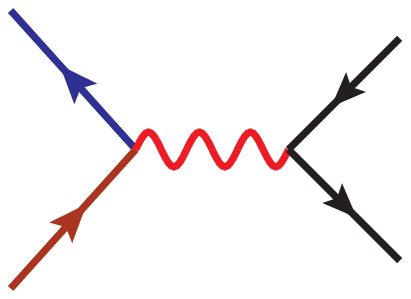}
  \\
  \parbox{0.6\textwidth}{%
    \caption{The Drell-Yan process in space.  The vertical lines
      with their dots signify Lorentz-contracted hadrons and their
      valence quarks. The star is the location of the hard collision. }
      \label{fig:DY}
    }
    &
    \parbox{0.3\textwidth}{%
      \caption{Parton-model hard scattering for Drell-Yan.}
      \label{fig:DY.hard.LO}
    }
  \end{tabular}
\end{figure}

Given the complications in QCD, it is useful to first recall the ideas
embodied in Drell and Yan's original model\cite{Drell:1970wh} for the
DY process, $AB\to l^+l^-X$, illustrated in Fig.\ \ref{fig:DY}.  Two
Lorentz-contracted hadrons collide at a high center-of-mass energy
$\sqrt{s}$.  The model has a short-distance hard collision with a
quark-antiquark annihilation through an electroweak boson, e.g.,
$q\bar{q} \to \gamma^* \to l^+l^-$, treated to lowest order.  The
measured transverse momentum $\T{q}$ of the lepton pair is the sum of
the transverse momenta of the annihilating partons, so that the
$\T{q}$ dependence of the DY cross section directly probes the
distributions of parton transverse momenta.

For the kinematic variables, I use light-front coordinates
$V=(V^+,V^-,\T{V})$, with the hadrons $A$ and $B$ moving in the $+z$ and
$-z$ directions.  I let $Q$ and $y=\frac12\ln\frac{q^+}{q^-}$ be the
invariant mass and rapidity of the lepton pair, and I define Bjorken
variables $x_A=Qe^y/\sqrt{s}$ and $x_B=Qe^{-y}/\sqrt{s}$.

The model's cross section differential in $q^\mu$ and in the lepton
angle is
\begin{equation}
\label{eq:DY.pm}
  \frac{ \diff{\sigma} }{ \diff[4]{q}\diff{\Omega} }
  ~\eqbad~ \frac{2}{s} \sum_j \int \diff[2]{\Tj{k}{A}} 
       ~f_{j/A}(x_A,\Tj{k}{A})
       ~f_{\bar{\jmath}/B}(x_B,\T{q}-\Tj{k}{A})
       ~\frac{ \diff{\hat{\sigma}_{\text{LO},j\bar{\jmath}}} }
             { \diff{\Omega} }.
\end{equation}
Here, $f_{j/H}(x,\T{k})$ is the TMD density of a parton of flavor $j$
in hadron $H$, with $x$ and $\T{k}$ being the parton's fractional
longitudinal momentum and transverse momentum.  The hard scattering
factor $\diff{\hat{\sigma}_{\text{LO},j\bar{\jmath}}}/\diff{\Omega}$ is the
lowest-order cross section for $q\bar{q} \to l^+l^-$, from the graph of
Fig.\ \ref{fig:DY.hard.LO}.  The question mark in Eq.\
(\ref{eq:DY.pm}) indicates that the formula is not fully correct in
QCD.

\section{Extension of the parton model to QCD}

Complications arise in correcting the parton model idea to apply to
real QCD. One is that the parton model intuition is natural in
coordinate space, Fig.\ \ref{fig:DY}, but calculations etc are
formulated in momentum space.  Thus a certain fuzziness occurs in
matching the two views.  Next, typical analyses use (all-orders)
perturbation theory, but there are clearly important non-perturbative
parts of QCD such as are accommodated in the parton densities; the
parton model ideas suggest that the perturbative analysis gives
properties of full QCD.  Finally, there are complications in QCD that
distort and modify the basic parton-model intuition.

\subsection{Space-time issues: interference in transverse coordinate
  space}

\begin{figure}
  \centering
  \(
   \psfrag{b}{\small $\T{b}$}
     \includegraphics[scale=0.45]{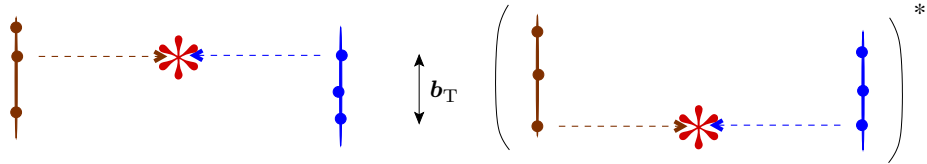}
  \)
  \caption{Amplitude times complex conjugate amplitude in coordinate space.}
  \label{fig:DY.bT}
\end{figure}

To obtain the transverse-momentum distribution of the cross section, there
is a Fourier transform over the \emph{difference} $\T{b}$ of
transverse positions of the hard scattering in the amplitude and its
complex conjugate: $\int \diff[2]{\T{b}} e^{i\T{q}\cdot\T{b}} \dots$,
as in Fig.\ \ref{fig:DY.bT}.  This gives interference between
scattering at different transverse positions, with a characteristic
wavelength proportional to $1/\Tsc{q}$.  In contrast the
resolution/wave-length for longitudinal distances is proportional to
$1/Q$.  The cross section integrated over all $\T{q}$ gives zero
transverse separation: $\T{b}=0$.

\subsection{Simplest candidate QFT translation}

\begin{figure}
  \centering
  \begin{minipage}{0.45\linewidth}
    \centering
    \includegraphics[scale=0.45]{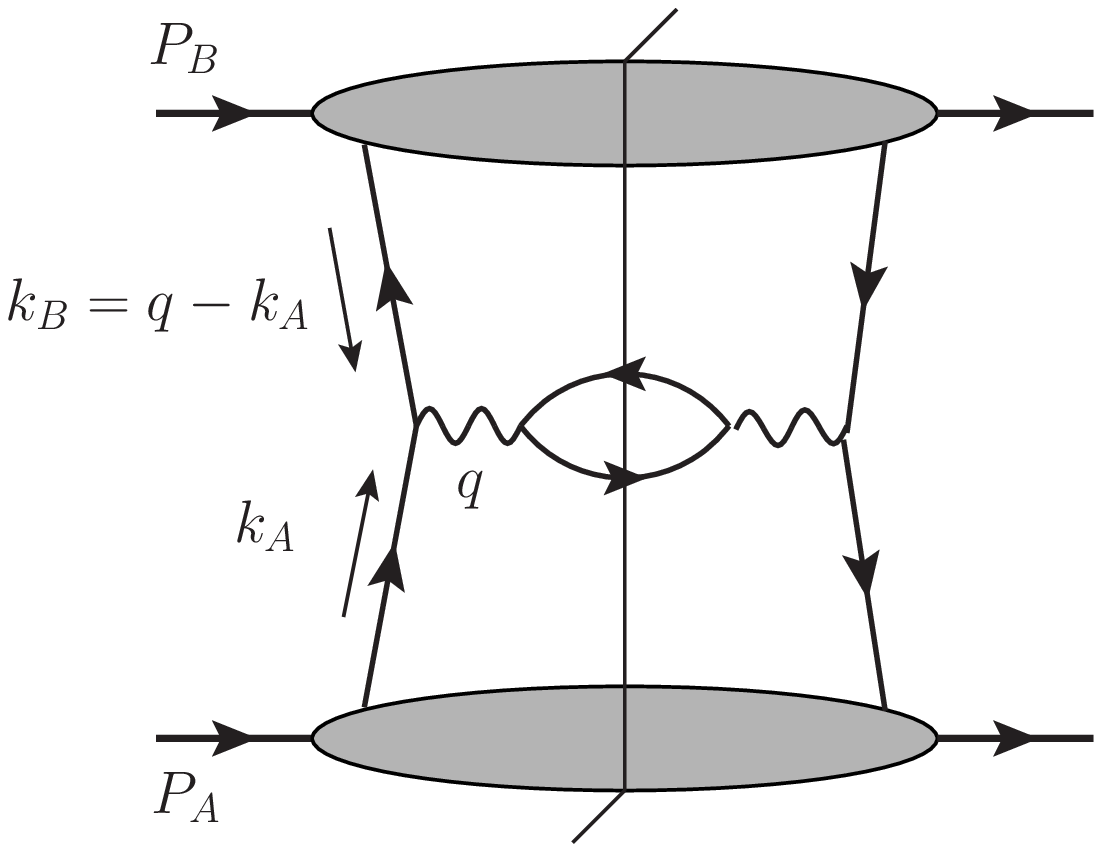}
    \caption{Parton model for DY process in terms of momentum space
      amplitudes.}
    \label{fig:DY.pm.graph}
  \end{minipage}
  \hfill
  \begin{minipage}{0.45\linewidth}
    \centering
    \includegraphics[scale=0.22]{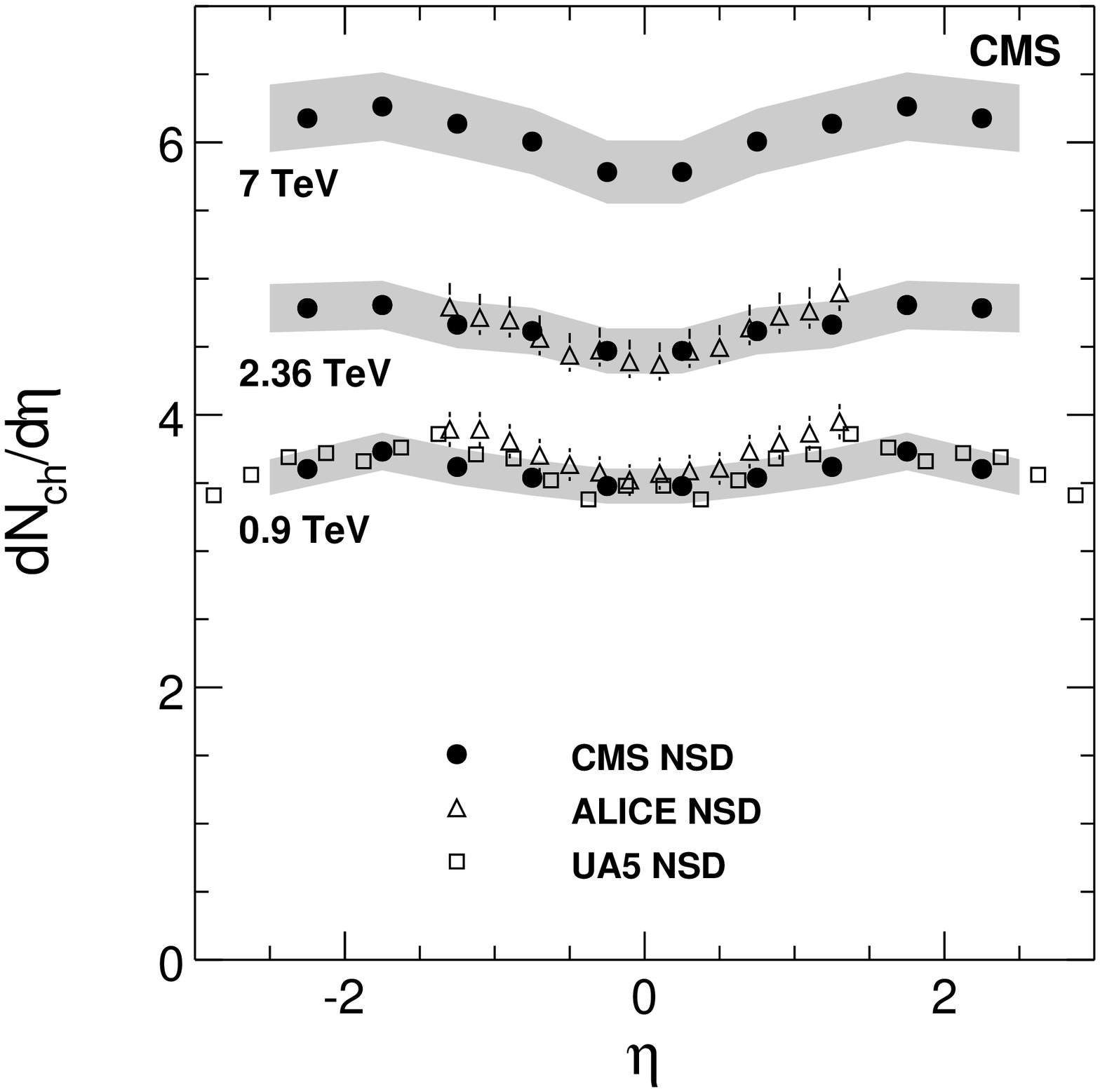}
    \caption{Hadron distribution in pseudo rapidity%
      \protect\cite{Khachatryan:2010us}.
    }
    \label{fig:eta.distrib}
  \end{minipage}
\end{figure}

In momentum-space terms, the parton model for DY uses just the
graphical structure of Fig.\ \ref{fig:DY.pm.graph}.  Attached to each
incoming hadron is a subgraph dominated by momenta collinear to the
hadron.  The only connecting lines are the single annihilating parton
on each side of the final state cut.  However, there are in reality
``spectator-spectator'' interactions that need to be examined, even
though their effects actually cancel in the inclusive cross section.
QCD has further non-trivial gluonic effects.

%

\subsection{Spectator-spectator interactions}

\begin{figure}
  \centering
    \begin{tabular}{cc}
      \VC{\includegraphics[scale=0.4]{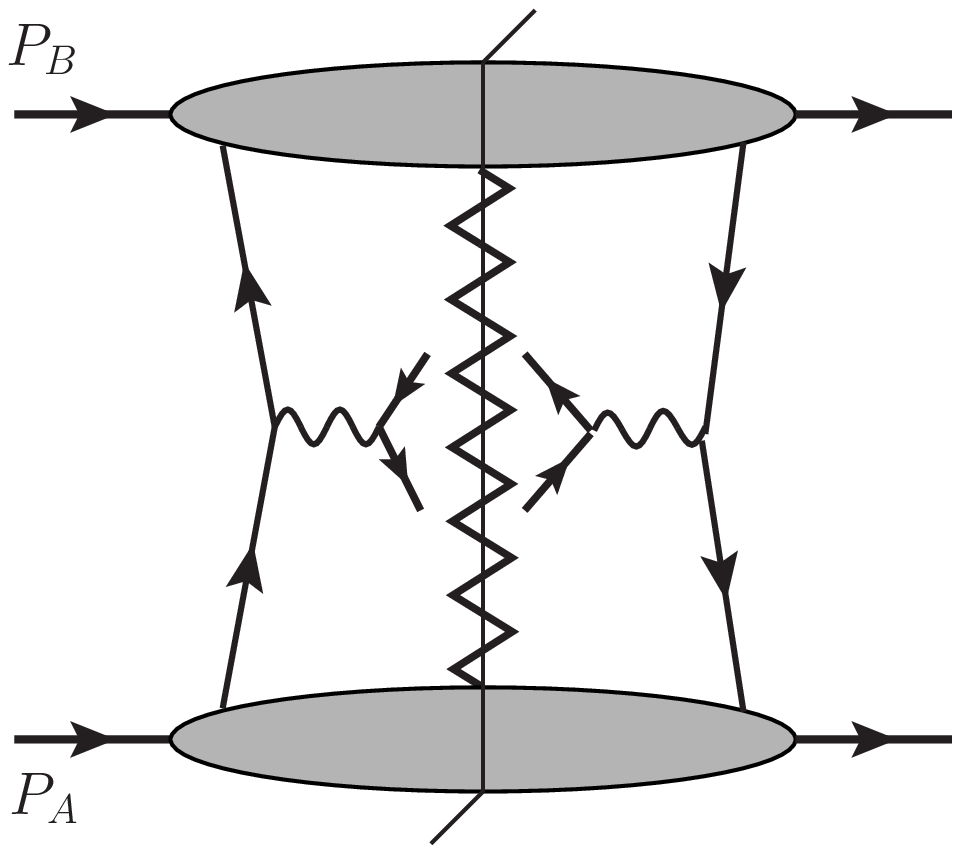}}
      &
      \VC{\includegraphics[scale=0.4]{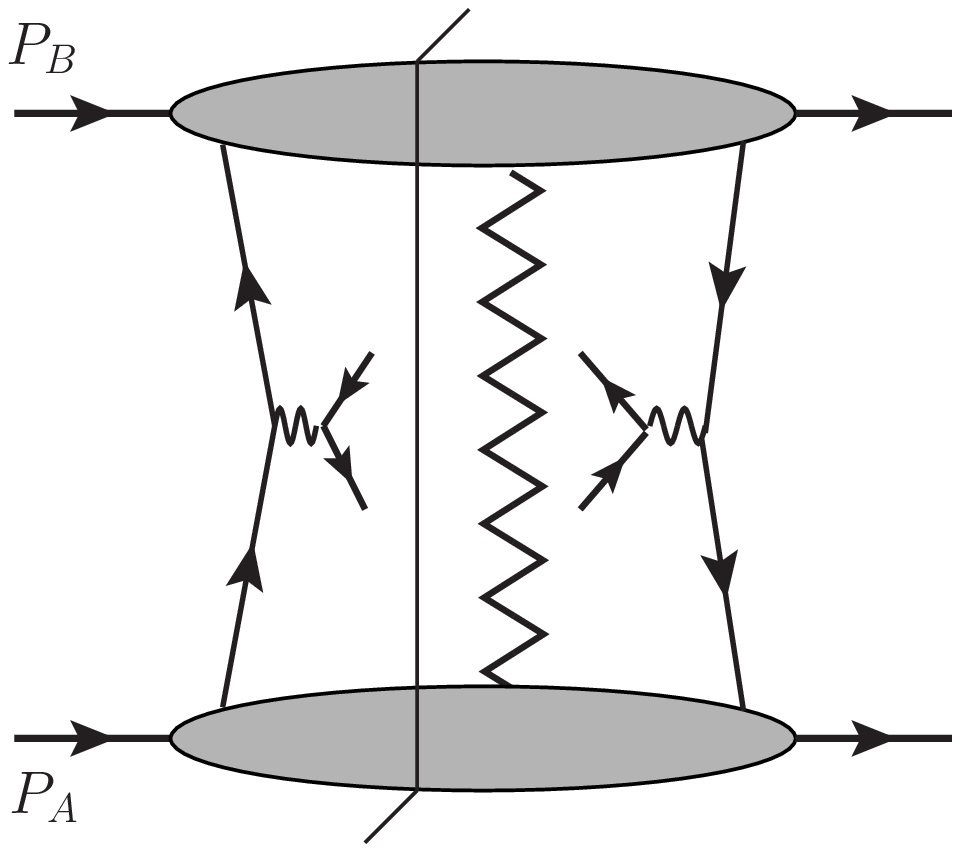}}
    \\
      (a) & (b)
    \end{tabular}
  \caption{Structure of simple graphs with final-state
    interactions. The zigzag line indicates a general graphical
    structure, often modeled as a sum over ladder graphs or as a
    pomeron.}
  \label{fig:fsi}
\end{figure}

Non-trivial spectator-spectator interactions must exist, because Fig.\
\ref{fig:DY.pm.graph} by itself gives colored particles in the
hadronic final states with a large rapidity gap.  In contrast the
final state in hadron-hadron collisions normally contains hadrons that
are distributed approximately uniformly in rapidity: Fig.\
\ref{fig:eta.distrib}.  In Fig.\ \ref{fig:fsi}, graphs of the form of
(a) can fill in the rapidity gap, while graphs like (b) make a
reduction in the rapidity-gap cross-section.

The sum over these interactions cancels\cite{DeTar:1974vx} to leading
power in the inclusive cross section.  In coordinate space, the
cancellation is more intuitive: These interactions happen outside past
light-cone of the hard scattering, i.e., too late to affect it.  The
cancellation only applies to the inclusive DY process: Cross sections
in which requirements are imposed on the hadronic final state are a
different matter\cite{DeTar:1974vx}.

\subsection{Kinematic region of gluons: 1-loop paradigm}

\begin{figure}
  \centering
  \begin{tabular}{p{0.45\textwidth}p{0.45\textwidth}}
    \VC{\includegraphics[scale=0.4]{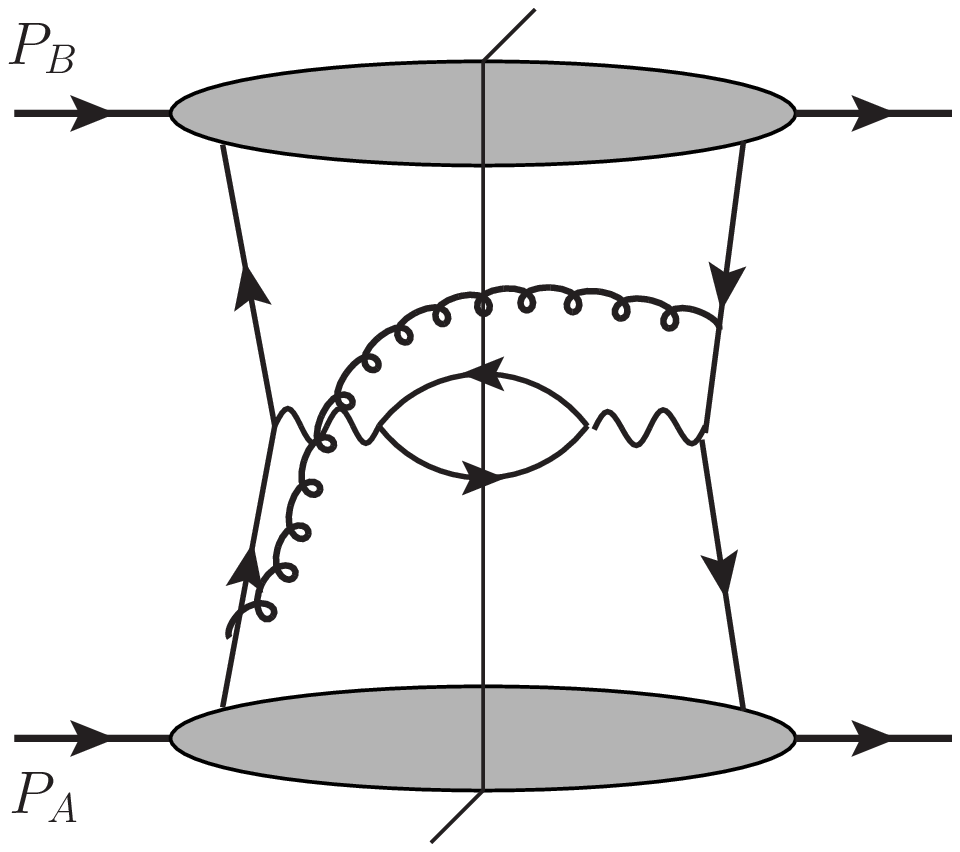}}
    &
      \psfrag{y}{\small $y_k$}
      \psfrag{yA}{\small $y_{p_A}$}
      \psfrag{yB}{\small $y_{p_B}$}
      \psfrag{ln kT/m}{\small $\ln(\Tsc{k}/m)$}
      \psfrag{k-~pB-}{\small \hspace*{-15mm}$k^-\sim p_B^-$}
      \psfrag{k+~pA+}{\small $k^+\sim p_A^+$}
      \psfrag{k-~pB-}{}
      \psfrag{k+~pA+}{}
    \VC{\includegraphics[scale=0.8]{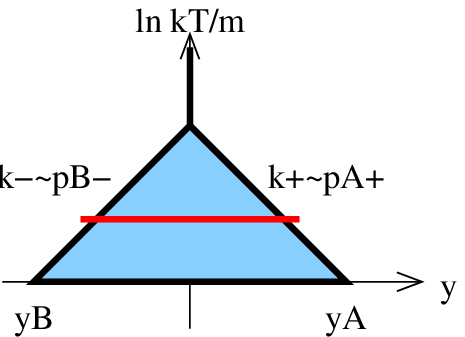}}
  \\
    \caption{One real gluon added to parton-model graph.}
    \label{fig:1g.graph}
    &
    \caption{Region of gluon momentum that is important in Fig.\
      \ref{fig:1g.graph}.}
    \label{fig:1g.region}
  \end{tabular}
\end{figure}

Extra terms that affect the TMD factorization formula in QCD are
illustrated by Fig.\ \ref{fig:1g.graph}, which gives a one-gluon
correction to the parton model.
The relevant kinematic range in the rapidity and the logarithmic
transverse momentum of the gluon is shown by the (red) horizontal line
in Fig.\ \ref{fig:1g.region}.  The gluon approximately has the
opposite transverse momentum to the DY pair, and it is emitted
essentially uniformly in rapidity, between kinematic limits.  The
dependence of the rapidity range on $\Tsc{q}$ is described by the
triangle.

Both left- and right-moving gluons are included in the important range
of momenta. To obtain factorization, the coupling of a gluon to an
oppositely moving quark is converted to a Wilson line vertex in the
operator definition of the TMD parton density.

\begin{figure}
   \centering
   {\includegraphics[scale=0.6]{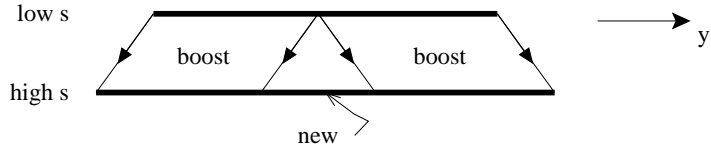}}
   \caption{Effect of boost on gluon in Fig.\
     \protect\ref{fig:1g.graph}.}
   \label{fig:boost}
\end{figure}

An important consequence, at all orders in gluon emission, is that the
$\T{q}$ distribution becomes energy dependent.  Consider Fig.\
\ref{fig:boost},
which illustrates the effect of increasing $s$ with $x_A$ and $x_B$
held fixed.  At the lower value of $s$, the range of gluon rapidity is
the upper thick line.  To get to higher $s$, the two incoming hadrons
are boosted, in opposite directions.  The annihilating quark and
antiquark are similarly boosted, because $x_A$ and $x_B$ are fixed.
Left- and right-moving gluons at the lower energy can be similarly
boosted, as indicated by the diagonal lines.  These effects alone give
unchanged transverse momentum.  But there is an extra region of gluon
emission, in the center at the bottom of the triangle.

The consequent broadening of the transverse-momentum distribution is a
definite prediction of QCD.  In the CSS-style formalism described
below, it appears as an energy dependence of the TMD parton densities.

\section{Full factorization}

The TMD factorization formula for the cross section in QCD is
\begin{align}
  &\hspace*{-5mm}
  \frac{ \diff{\sigma} }{ \diff[4]{q}\diff{\Omega} }
\nonumber\\
  ={}&  \frac{2}{s} \sum_j  
        \frac{ \diff{\hat{\sigma}_{j\bar{\jmath}}}(Q,\mu,g(\mu)) }
             { \diff{\Omega} }
        \int \diff[2]{\T{b}}
        ~ e^{i\Tj{q}{h}\cdot \T{b} }
        ~ \tilde{f}_{j/A}(x_A,\T{b};\zeta_A,\mu) 
        ~ \tilde{f}_{\bar{\jmath}/B}(x_B,\T{b};\zeta_B,\mu)
\nonumber\\
    & + \mbox{poln.\ terms} 
      + \mbox{high-$\Tsc{q}$ term}
      + \mbox{power-suppressed}
\end{align}
where $\mu$ is a renormalization scale, and each $\zeta$ is (up to
power corrections) $(2 \times \text{corresponding parton energy})^2$,
with $\zeta_A\zeta_B=Q^4$, e.g., $\zeta_A=\zeta_B=Q^2$.
Compared with the parton model formula, (\ref{eq:DY.pm}):
\begin{enumerate}
\item The hard scattering, $\diff{\hat{\sigma}}$ includes higher-order
  perturbative QCD corrections.
\item The TMD parton densities depend on two auxiliary parameters
  $\zeta$ and $\mu$.
\item Convolution in $\T{q}$ is replaced by multiplication in $\T{b}$.
\item There are similar terms involving polarization-dependence; these
  are extensively discussed elsewhere in the proceedings of this
  workshop.
\item TMD factorization is accurate when $\Tsc{q} \ll Q$.  A added
  correction or matching term using collinear factorization at high
  $\Tsc{q}$ gives a formula accurate for all $\Tsc{q}$.
\end{enumerate}

\subsection{Evolution, etc for TMD pdfs}
\label{sec:evol}

Evolution equations for the TMD parton densities are
\begin{align}
\label{eq:CSS}
  \frac{ \partial \ln \tilde{f}_{f/H}(x,\Tsc{b}; \zeta; \mu) }
       { \partial \ln \sqrt{\zeta} }
  &= 
  \tilde{K}(\Tsc{b};\mu),
\\
\label{eq:TMD.pdf.RG}
  \frac{ \diff{ \ln \tilde{f}_{f/H}(x,\Tsc{b};\zeta;\mu) }}
       { \diff{\ln \mu} }
  &= \gamma_f( g(\mu); 1 )
      - \frac12 \gamma_K(g(\mu)) \ln \frac{ \zeta }{ \mu^2 },
\intertext{with provably only a single logarithm of
    $\zeta/\mu^2$ on the right-hand-side of
    (\ref{eq:TMD.pdf.RG}).  The kernel $\tilde{K}$ of the CSS
    equation (\ref{eq:CSS}) obeys}
\label{eq:K.RG}
  \frac{ \diff{\tilde{K}} }{ \diff{\ln \mu } }
  &= -\gamma_K\xleft(g(\mu)\right).
\end{align}
At small-$\Tsc{b}$, the TMD parton densities have a generalized
operator-product expansion in terms of ordinary parton densities:
\begin{equation}
\label{eq:small.b}
  \tilde{f}_{f/H}(x,\Tsc{b};\zeta;\mu) 
  = \sum_j \int_{x-}^{1+} \frac{ \diff{\hat{x}} }{ \hat{x} }
       \,\tilde{C}_{f/j}\xleft( x/\hat{x},\Tsc{b};\zeta,\mu,g(\mu) \right)
       \, f_{j/H}(\hat{x};\mu)
~+~ O\xleft[(m\Tsc{b})^p \right],
\end{equation}
where the coefficient functions $\tilde{C}$ are perturbatively calculable.

\subsection{Exploit factorization, and evolution}

The above equations can be exploited to give predictive power to the
formalism. 
\begin{itemize}
\item Evolution can be used to remove large logarithms in quantities
  that have an intrinsically large momentum scale.\footnote{%
    Then in an exact analytic solution like (\ref{eq:sol2}) below, one
    can usefully replace quantities like $\diff{\hat{\sigma}}$, $\gamma_K$, etc
    by fixed-order perturbative approximations, with controllable
    errors for the cross section itself.  This procedure is to be
    distinguished from the related but weaker method of resummation,
    when resummation is interpreted to mean that a fixed-coupling
    expansion of relevant factors in the cross section is made, and a
    particular set of terms (e.g., leading logarithmic terms) is
    retained.  }
\item Intrinsically non-perturbative parts are in the large $\Tsc{b}$
  behavior of the TMD pdfs and of $\tilde{K}$, and are determined
  by fits or by non-perturbative calculations/modeling.
\end{itemize}

\subsection{Solutions}

I will present two solutions of the equations.  

\subsubsection{One solution: Factorization with fixed TMD pdfs}

In the first, the TMD parton densities appear with fixed scales.  This
is the closest form to the parton model formula (\ref{eq:DY.pm}),
where the parton densities can be treated as intrinsic and universal
properties of QCD.  The renormalization scale in the hard scattering
is chosen of order $Q$, so that it is usefully estimated by a
perturbative calculation at low-order.
\begin{align}
\label{eq:sol1}
  \frac{ \diff{\sigma} }{ \diff[4]{q}\diff{\Omega} } 
  ={}&  \frac{2}{s} \sum_j  
    \frac{ \diff{\hat{\sigma}_{j\bar{\jmath}}}(Q,\muQ,g(\mu)) }{ \diff{\Omega} }
    \int \diff[2]{\T{b}}
    e^{i\Tj{q}{h}\cdot \T{b} }
    \times 
\nonumber\\&
    \times 
    \tilde{f}_{j/A}\big( x_A, \T{b}; m^2,\mu_0 \bigr)
    \,
    \tilde{f}_{\bar{\jmath}/B}\big( x_B, \T{b}; m^2,\mu_0 \bigr)
\nonumber\\&
  \times
  \left( \frac{ Q^2 }{ m^2 } \right)^{\tilde{K}(\Tsc{b};\mu_0) }
  \times
  \exp\left\{   
           \int_{\mu_0}^{\muQ}  \frac{ \diff{\mu'} }{ \mu' }
           \biggl[ 2 \gamma(g(\mu'); 1) 
                 - \ln\frac{Q^2}{ (\mu')^2 } \gamma_K(g(\mu'))
           \biggr]
  \right\}
\nonumber\\&
+ \mbox{polarized terms}
+ \mbox{large $\Tscj{q}{h}$ correction, $Y$}
+ \mbox{p.s.c.}
\end{align}
where $\muQ \propto Q$, while $m$ and $\mu_0$ are fixed scales.  The
scale $\mu_0$ should be in a perturbative region, so that anomalous
dimensions can be treated perturbatively.

In addition to the TMD parton densities, there is the evolution kernel
$\tilde{K}$ (at a fixed scale).  This gives a power-law dependence on
$Q$, with the power depending on $\Tsc{b}$.  It therefore determines
the $Q$ dependence of the shape of the $\Tsc{q}$ distribution.

\subsubsection{Another solution: For maximum perturbative content}

The second solution, as presented by CSS\cite{Collins:1984kg}, has
maximum perturbative content: The small-$\Tsc{b}$ expansion
(\ref{eq:small.b}) is used when possible, and evolution is applied to
its coefficients and to $\tilde{K}$ to remove large logarithms.
Remaining non-perturbative content is parameterized by functions to be
fit to data.
\begin{align}
\label{eq:sol2}
  &\frac{ \diff{\sigma} }{ \diff[4]{q}\diff{\Omega} } 
  =  \frac{2}{s}   \sum_{j,j_A,j_B}
        \frac{ \diff{\hat{\sigma}_{j\bar{\jmath}}}(Q,\mu_Q,g(\mu_Q)) }{ \diff{\Omega} }
        \int \frac{ \diff[2]{\T{b}} }{ (2\pi)^2 }  e^{i\Tj{q}{h}\cdot \T{b} }
\nonumber\\& \times
  e^{-g_{j/A}(x_A,\Tsc{b}) }
  \int_{x_A}^1 \frac{ \diff{\hat{x}_A} }{ \hat{x}_A }
       f_{j_A/A}(\hat{x}_A;\mub) 
  ~ \tilde{C}_{j/j_A}\xleft( \frac{x_A}{\hat{x}_A},\bstarsc; \mub^2, \mub, g(\mub) \right)
\nonumber\\& \times
  e^{ -g_{\bar{\jmath}/B}(x_B,\Tsc{b})}
  \int_{x_B}^1 \frac{ \diff{\hat{x}_B} }{ \hat{x}_B }
       f_{j_B/B}(\hat{x}_B;\mub) 
  ~ \tilde{C}_{\bar{\jmath}/j_B}\xleft( \frac{x_B}{\hat{x}_B},\bstarsc; \mub^2, \mub, g(\mub) \right)
\nonumber\\& \times  
  \left( \frac{ Q^2 }{ m^2 } \right) ^ {- g_K(\Tsc{b})}
  \left( \frac{ Q^2 }{ \mub^2 } \right) ^ {\tilde{K}(\bstarsc;\mub)}
  \exp\xleft\{ 
       \int_{\mub}^{\muQ}  \frac{ \diff{\mu'} }{ \mu' }
          \left[ 2 \gamma(g(\mu'); 1) 
                 - \ln\frac{Q^2}{ (\mu')^2 } \gamma_K(g(\mu'))
          \right]
  \right\}
\nonumber\\&
+ \mbox{polarized terms}
+ \mbox{large-$\Tscj{q}{h}$ correction, $Y$}
+ \mbox{p.s.c.}
\end{align}
Here, CSS chose $\bstar = \ifrac{ \T{b} }{ \sqrt{ 1 + \Tsc{b}^2/b_{\rm
      max}^2} }$, with $b_{\rm max}$ a constant chosen so that
$\bstar$ never goes too far beyond the perturbative region.  (The
appropriateness of choices of $b_{\rm max}$ is under active discussion
currently, as can be seen in several other contributions --- e.g.,
those of Boer, Idilbi, Prokudin, and Yuan.)  The scale $\mub$ is
proportional to $1/\bstarsc(\Tsc{b})$.  Non-perturbative $\Tsc{b}$
dependence is contained in the functions $g_{j/A}(x_A,\Tsc{b})$,
$g_{\bar{\jmath}/B}(x_B,\Tsc{b})$, and $g_K(\Tsc{b})$.

\subsection{Evolution in $\Tsc{q}$ v.\ $\Tsc{b}$}

Fig.\ \ref{fig:xsect}
\begin{figure}
  \newcommand\thisscale{0.2}
  \begin{tabular}{c@{\hspace*{1cm}}c@{\hspace*{1cm}}c}
   Low $Q$, $\sqrt{s}$ 
   & \VC{\includegraphics[scale=\thisscale]{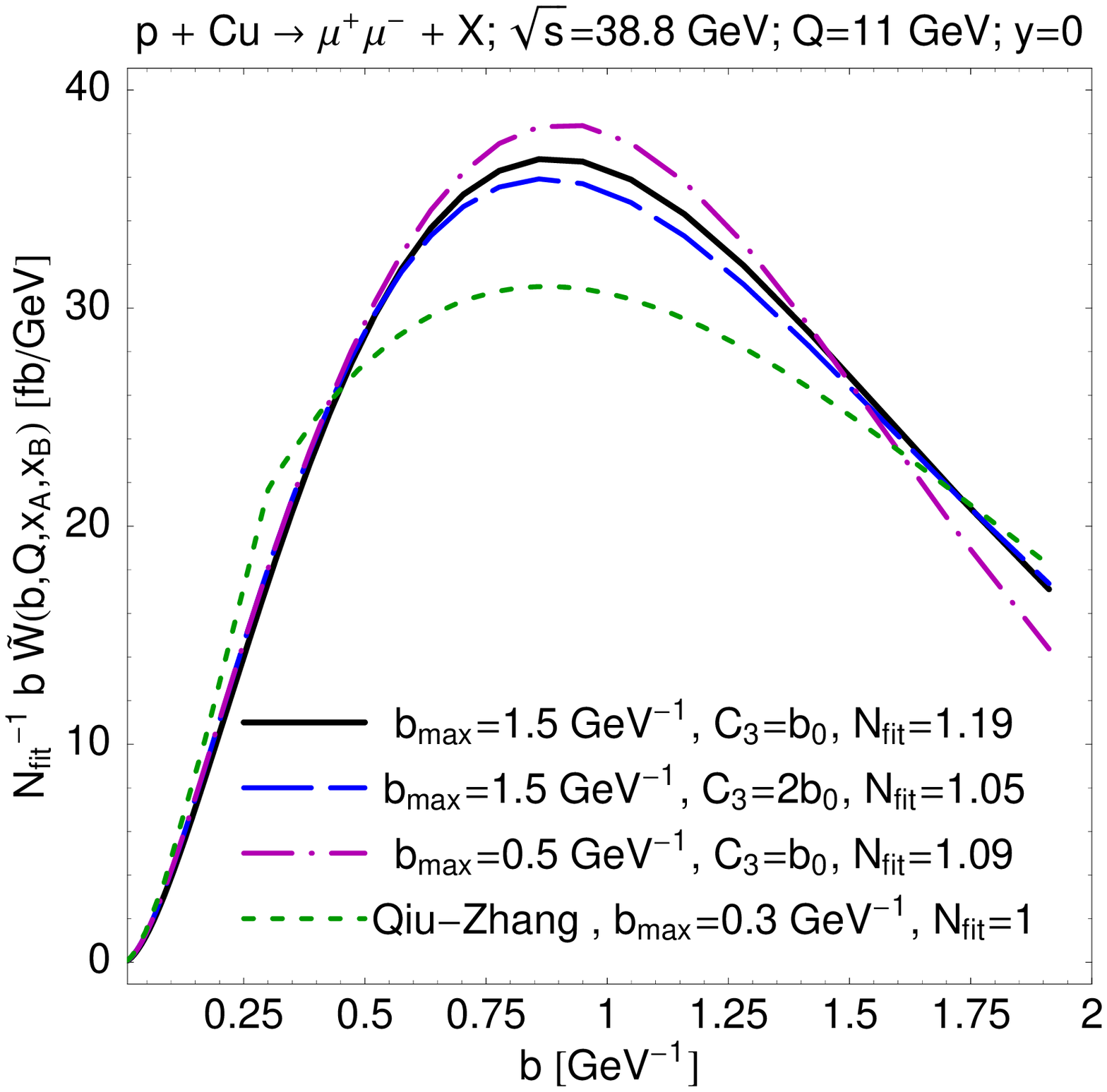}}
   & \VC{\includegraphics[scale=\thisscale]{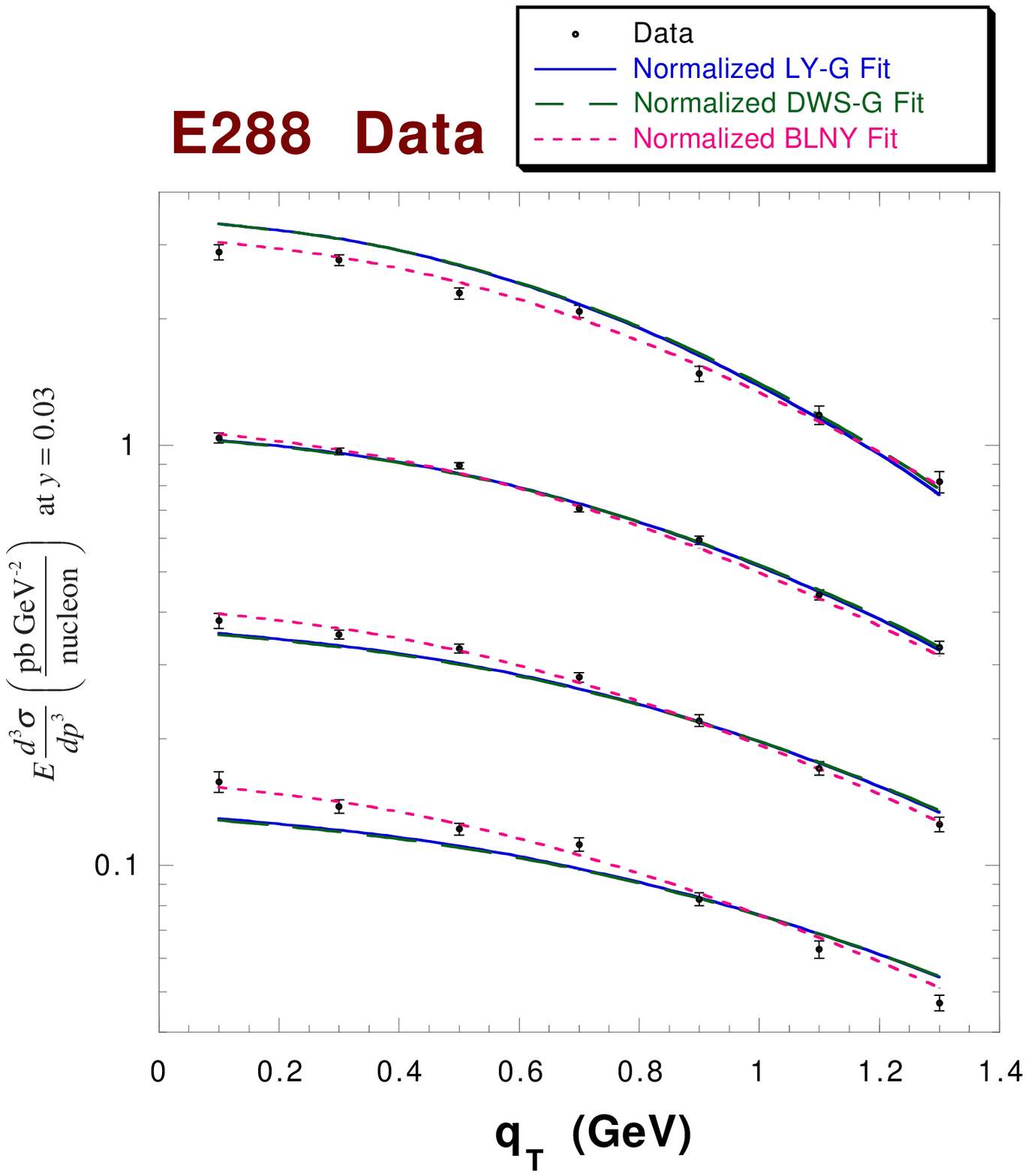}}
  \\
   High $Q$, $\sqrt{s}$ 
   & \VC{\includegraphics[scale=\thisscale]{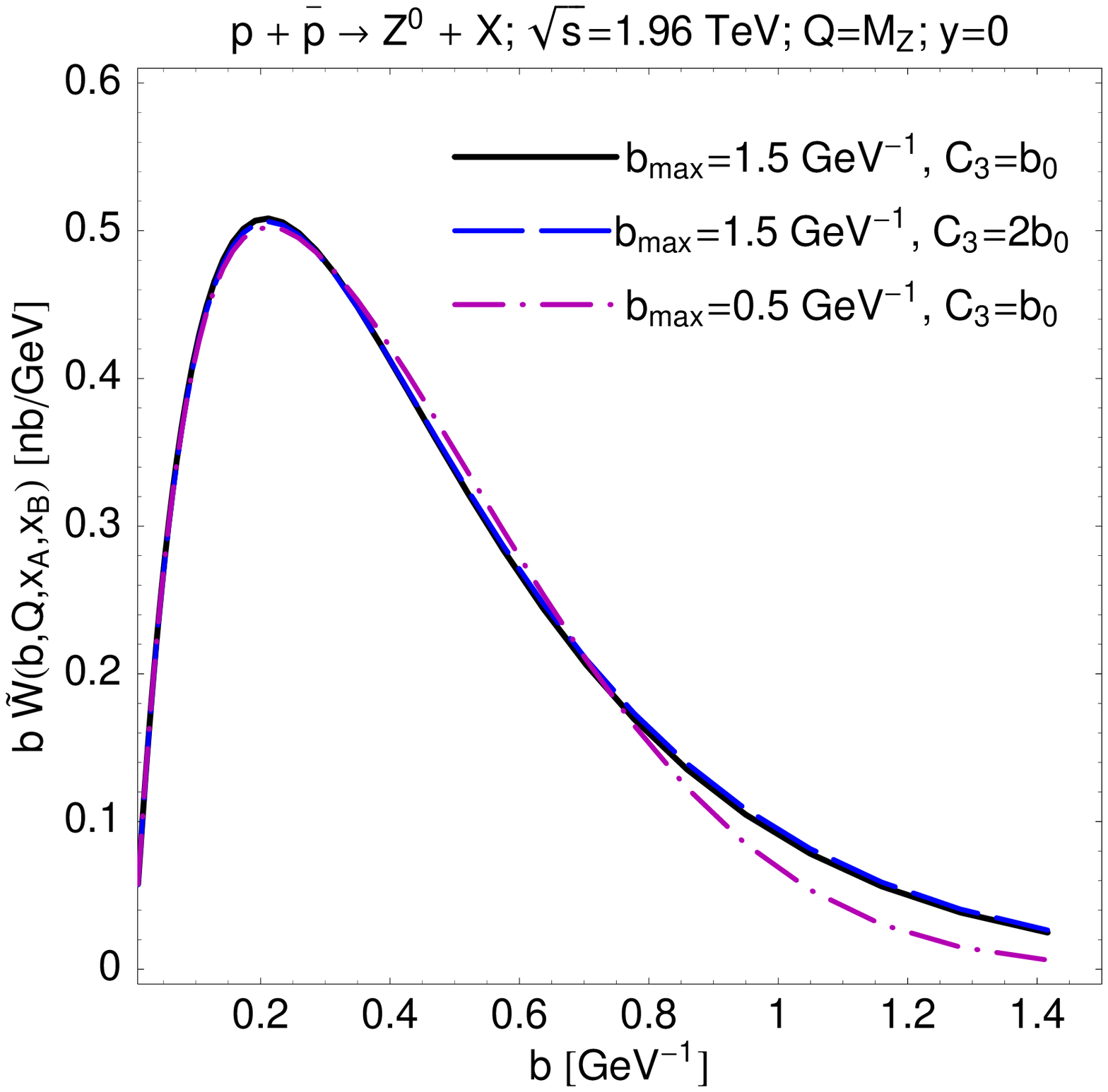}}
   & \VC{\includegraphics[scale=\thisscale]{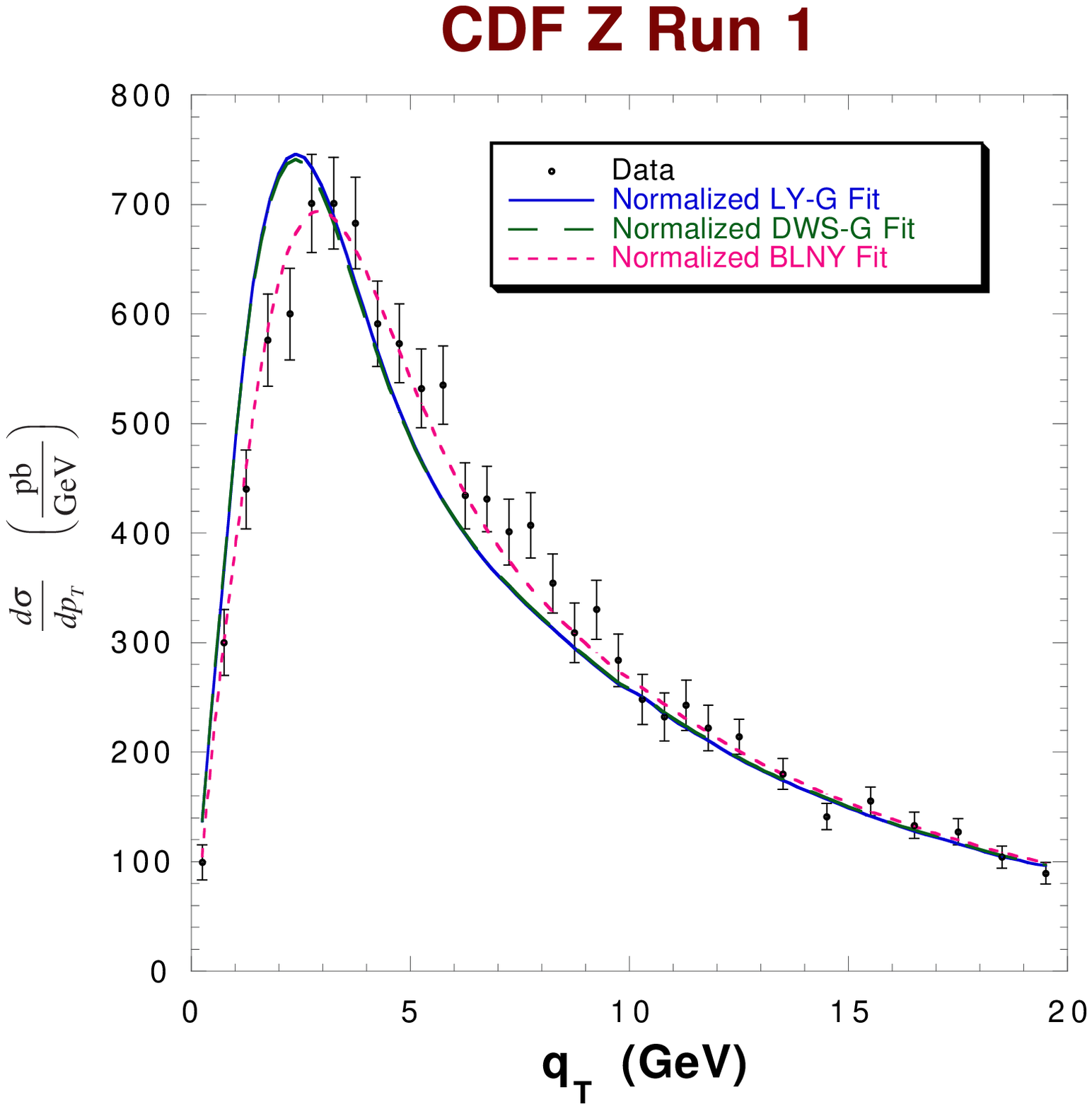}}
  \end{tabular}
  \caption{On the left: Integrand\protect\cite{Konychev:2005iy} in
    $\Tsc{b}$ space at low values $Q$ and $s$ and at high values.  On
    the right: corresponding plots\protect\cite{Landry:2002ix} for the
    $\Tsc{q}$ dependence of the cross section.  The plot with CDF data
    has a zero at $\Tsc{q}=0$
    because the data are given for $\diff{\sigma}/\diff{\Tsc{q}}$
    instead of $\diff{\sigma}/\diff{\Tsc{q}^2}$.  }
  \label{fig:xsect}
\end{figure}
shows results for evolution both in $\Tsc{b}$ space and after Fourier
transformation to transverse momentum.  As $Q$ increases, the
high-$\Tsc{b}$ tail is strongly suppressed, so a perturbatively-based
calculation of the transverse momentum distribution becomes accurate,
with the remaining non-perturbative quantitative information being in
the ordinary integrated parton densities. The situation is different
at relatively low $Q$.  Correspondingly the $\Tsc{q}$ distribution
broadens with energy.

\section{What form for large $\Tsc{b}$?}

In some standard fits\cite{Konychev:2005iy,Landry:2002ix}, a Gaussian
distribution is assumed for the intrinsic transverse-momentum
functions, $e^{-\text{const} \times \Tsc{b}^2}$ at large $\Tsc{b}$.
Correspondingly $\tilde{K}$ is assumed to be quadratic in $\Tsc{b}$ at
large $\Tsc{b}$, which gives an energy dependent Gaussian in the
factorization formula: $e^{-\tilde{K}\ln Q^2} \sim e^{-\text{const}
  \times \Tsc{b}^2\ln Q^2}$.  The coefficients in the Gaussian
exponent are substantially non-zero according to the fits in Refs.\
\refcite{Konychev:2005iy,Landry:2002ix}.

These assumptions should be questioned\cite{Schweitzer:2012hh}, since
Euclidean correlation functions in QFT are usually exponential,
$e^{-m\Tsc{b}}$, not Gaussian at large distances.  Furthermore,
Schweitzer, Strikman and Weiss\cite{Schweitzer:2012hh} argue that
there are two relevant non-perturbative scales: a chiral scale
$\unit[0.3]{fm} = \unit[1.5]{GeV^{-1}}$ and a confinement scale =
$\unit[1]{fm} = \unit[5]{GeV^{-1}}$, each with characteristic effects
on the sea and valence quark densities.  If the scale $m$ is
$Q$-independent, then the functions $g_{j/H}(x,\Tsc{b})$ in Eq.\
(\ref{eq:sol2}) are linear at large $\Tsc{b}$, while the function
$g_K(\Tsc{b})$ goes to a constant.

In principle the value of $b_{\rm max}$ is irrelevant; any change is
compensated by a change in the functional form of the non-perturbative
functions.  In practice, it is probably preferable to use the
information in perturbative calculations as much as possible, so that
one should prefer larger values of $b_{\rm max}$.  From Ref.\
\refcite{Schweitzer:2012hh}, it is reasonable that Landry et
al's\cite{Landry:2002ix} $b_{\rm max} = \unit[0.5]{GeV^{-1}} =
\unit[0.1]{fm}$ is too low, and that Konychev and
Nadolsky's\cite{Konychev:2005iy} $\unit[1.5]{GeV^{-1}} =
\unit[0.3]{fm}$ is better.  Perhaps an even larger value is sensible.
In any case it is probable that the large quadratic terms in the
Landry et al fits\cite{Landry:2002ix} are mostly reproducing the
results of perturbation theory for $\Tsc{b}$ between about
$\unit[0.1]{fm}$ and $\unit[0.3]{fm}$, rather than giving the true
asymptotic behavior at large $\Tsc{b}$.

In view of the above, I suggest retrying fits with the following forms
at large $\Tsc{b}$:
\begin{itemize}
\item $e^{-\text{const} \times \Tsc{b}}$ in TMD parton densities, with
  different constants for sea and valence quarks.  A possibly useful
  parameterization for the $e^{ -g_{j/H}(x,\Tsc{b})}$ factor in TMD
  densities is $e^{ -m \left( \sqrt{\Tsc{b}^2+b_0^2} -b_0 \right) }$,
  which is exponential at asymptotically large $\Tsc{b}$, but
  approximately Gaussian at relatively small $\Tsc{b}$.
\item $e^{-\tilde{K}\ln Q^2} \to e^{-\text{const} \times \ln Q^2}$ at large
  $\Tsc{b}$ in the evolution factor.
\end{itemize}

\section{Predictions, issues}

In principle, the TMD factorization framework is highly predictive
(and hence testable).  Basically, one can fit the non-perturbative
$\Tsc{b}$-dependence of TMD functions at low energy.  The dependence
of one process on $Q$, for a limited range of $Q$ with fixed $x$ is
sufficient to fit the non-perturbative $\tilde{K}$.  Then everything
else is predicted.  Using polarized TMD parton densities and
fragmentation functions does not need new values of $\tilde{K}$.
There is also the predicted sign reversal of naively T-odd TMD parton
densities (the Sivers function, etc) between DY and SIDIS.

The standard processes are (a) DY, sensitive to TMD parton densities
in certain combinations; (b) SIDIS, sensitive to both TMD parton
densities and fragmentation functions, in many flavor combinations;
(c) $e^+e^- \to \text{back-to-back hadrons}$, sensitive to TMD
fragmentation functions, including flavor dependence.

Given the recent extra data, especially at relatively low $Q$, and
given the issues about the functional form of the non-perturbative
$\Tsc{b}$ dependence, it is important to update global fits beyond
Refs.\ \refcite{Konychev:2005iy,Landry:2002ix}.
In addition, as can be seen from other contributions to this workshop,
there is an urgent need to reconcile treatments. 

Finally, note the predicted \emph{violation of TMD factorization in
  hadro-production of hadrons} --- see, for example, the contribution
of Rogers for new work.  Much work here is needed; it is an important
source of new phenomena in QCD.  Fits to data where TMD factorization
is valid are important in quantitatively assessing factorization
breaking elsewhere.

\section*{Acknowledgments}

This work was supported by the U.S. D.O.E. under grant number
DE-SC0008745.

\bibliographystyle{ws}
\providecommand\BIBvol[1]{\textbf{#1}}
\bibliography{jcc}

\end{document}